\tolerance=2500
\headline={\ifnum\pageno=1 \hfil
                      \else                         \hss\tenrm
Collier and Stubbs: Lunar Dust Generated Neutral Solar Wind
\hss\tenrm\folio\voffset=1.66in\fi}
\footline={\ifnum\pageno=1 \hss\tenrm\folio\hss
                    \else                          \hfil\fi}
\vskip 3.0in
\hoffset=-0.25in
\voffset=0.25in
\vsize=9.50in
\hsize=6.75in
\tenrm
\vskip 0.25in
\noindent
{\tenbf Neutral Solar Wind Generated by Lunar Exospheric Dust at the Terminator}
\vskip .10in
\noindent
\vskip 10pt \noindent
Michael R. Collier${}^1$ and Timothy J. Stubbs${}^{1,2}$ 
\vskip 0.0in \noindent
{\sevenrm 1. NASA/Goddard Space Flight Center, Greenbelt, Maryland, USA}
\vskip 0.0in \noindent
{\sevenrm 2. Goddard Earth Science and Technology Center, Univ. of Maryland, Baltimore County, USA}
\vskip 10pt
\noindent
\vskip 0pt \noindent
date: 30 August 2008 -- version: 3.2 
\vskip 0pt \noindent
\lineskiplimit=0in
\baselineskip=18pt
\lineskip=0in
\vskip .10in
{\noindent
{\tenbf Abstract.} 
We calculate the flux of neutral solar wind observed on the lunar surface at the terminator due to solar wind protons penetrating exospheric dust grains with (1) radii greater than 0.1$\mu$m and (2) radii greater than 0.01$\mu$m. For grains with radii larger than 0.1$\mu$m, the ratio of the neutral solar wind flux produced by exospheric dust to the incident ionized solar wind flux is estimated to be $\sim$10${}^{-4}$-10${}^{-3}$ for solar wind speeds in excess of 800~km/s, 
but much lower ($<$10${}^{-5}$) at average to slow solar wind speeds. However, when the smaller grain sizes are considered, this ratio is estimated to be $\ge$10${}^{-5}$ at all speeds, and at speeds in excess of 700~km/s reaches $\sim$10${}^{-3}$. These neutral solar wind fluxes are easily measurable with current low energy neutral atom instrumentation. Observations of neutral solar wind from the surface of the Moon would provide independent information on the distribution of very small dust grains in the lunar exosphere that would complement and constrain optical measurements at ultraviolet and visible 
wavelengths. 

%
%
%
\baselineskip=20pt
\vskip 20pt
\noindent{\tenbf 1. Introduction}

\nobreak

\vskip 5pt

\nobreak

Although the vast bulk of solar wind hydrogen is ionized, a small fraction 
of the solar wind observed at 1~AU and beyond in the inner heliosphere is neutral. The neutralization likely occurs either through charge exchange with other neutral atoms [Collier et al., 2003]; for example, exospheric or interstellar neutrals, or by penetration of small dust grains [e.g. Wimmer-Schweingruber and Bochsler, 2003].

Despite neutral solar wind having been discussed for many years [Fahr, 1968; Holzer, 1977] it has only recently been observed by the Low Energy Neutral Atom (LENA) imager at Earth [Collier et al., 2001] and by the Analyzer of Space Plasma and EneRgetic Atoms (ASPERA-3) at Mars [Brinkfeldt et al., 2006]. At Earth, neutral solar wind generated by the charge-exchange interaction of solar wind protons with the exosphere has provided remote observations of changes in the location of the magnetosphere and high latitude reconnection points [e.g. Taguchi et al., 2005, 2006; Collier et al., 2005]. The difficulty observing neutral solar wind arises from both technical challenges associated with detecting low energy neutral atoms and the potential high background levels due to stray light that results from looking in a direction close to the Sun. In LENA's case, these problems were mitigated by conversion surface technology, a light-tight design with a hemispherical analyzer, and a time-of-flight unit to differentiate particles from photons. In ASPERA's case, the observations at Mars were taken during a period when Mars Express was entering eclipse after solar occultation but before Mars blocked the neutral solar wind signal.

Future neutral solar wind observations may be made on ESA's Solar Orbiter mission which could fly the first dedicated neutral solar wind analyzer [J. Hsieh, private communication, 2007].

Although the solar wind contains a neutral component arising from many different sources, we focus here on the locally-generated lunar neutral solar wind as observed by a hypothetical neutral atom instrument similar to those flown on previous missions and deployed on the lunar surface in much the same way as the 
Apollo Lunar Surface Experiments Package
(ALSEP) and the proposed lunar sortie science missions.

\vskip 20pt
\noindent{\tenbf 2. Lunar Dust}

\nobreak

\vskip 5pt

\nobreak
Without exception, the astronauts who walked on the surface of the Moon
confronted problems due to lunar dust: it adhered to clothing and equipment, it reduced visibility, and it caused difficulty breathing [e.g., see Stubbs et al., 2007a and references therein].
In fact, ``the invasive nature of lunar dust represents a more challenging engineering design issue, as well as a health issue for [lunar] settlers, than
does radiation" [Schmitt, 2006].
Furthermore, lunar dust may pose acute toxicity risks to astronauts [Liu et al., 2007; Park et al., 2006]. 

Evidence from Surveyor vidicon images of horizon glow [Rennilson and Criswell, 1974], as well as excess brightness in photographs of the solar corona taken by Apollo astronauts just inside the Moon's shadow [McCoy, 1976] and Apollo astronaut observations of horizon glow and ``streamers" immediately prior to orbital sunrise [McCoy and Criswell, 1974], all suggest that a substantial population of exospheric dust exists in the terminator region of the Moon. 
Rennilson and Criswell [1974] concluded that $\sim$10${}^{7}$ more particles per unit time must be ejected than can be achieved by meteoritic impacts in order
to produce a dust cloud that could scatter enough sunlight to create the horizon glow observed by the Surveyor landers.

Solar photoelectron emission on the dayside tends to drive the lunar surface 
positive, typically about +10~V, while solar wind electrons incident on the nightside drives the surface negative, typically about -100~V [Freeman and Ibrahim, 1975; Stubbs et al., 2007b,c], although occasionally regions of the lunar surface can charge up to
a few kilovolts negative [Halekas et al., 2007]. It is believed that differential charging of the lunar surface in the terminator region 
can result in strong local electric fields that are able to eject charged
dust into the exosphere [e.g., Criswell and De, 1977; De and Criswell, 1977; Borisov and Mall, 2006; Wang et al., 2007].

From the limited observations of exospheric dust phenomena, it appears
that large grains (for our purpose) of $\sim$5-6~$\mu$m radius are ``levitated" to heights of $\sim$10-30~cm [Rennilson and Criswell, 1974] while the smaller grains, $\sim$0.1~$\mu$m, which are more relevant to neutral solar wind formation, are ``lofted" to heights in excess
of $\sim$100~km [McCoy, 1976]. It is estimated that the exospheric dust observed
by the Apollo astronauts had a scale height of $\sim$10-20~km [Zook and McCoy, 1991]. Ultraviolet horizon glow observations by Page and Carruthers [1978]
on the lunar dayside could be interpreted as being due to dust with radii as small as 0.01~$\mu$m. Recently, the Apollo samples have revealed evidence for lunar dust grains as small as 0.01~$\mu$m in diameter [Taylor, 2007; Liu et al., 2007].

The electric field configuration near the terminator region appears to drive the electrostatic transport of charged dust [Farrell et al., 2007; Stubbs et al., 2006, 2007c]. The most compelling evidence for this came from the Lunar Ejecta and Meteorites (LEAM) experiment deployed on the surface of the Moon by Apollo~17 astronauts [Berg et al., 1976]. LEAM was designed to detect very small fluxes of hypervelocity ($>$10~km/s) impacts from interplanetary and
interstellar dust, but instead detected significantly larger fluxes of slower moving ($\sim$100~m/s) highly charged dust grains of lunar origin [Berg et al., 1976; Colwell et al., 2007]. 
The limited observations of the lunar ``dust-plasma" environment suggest that it
is most active at the terminator, therefore we focus our attention on dust-created neutral solar wind near this region. In addition, this is also the region where this environment will most likely pose the greatest 
hazard to human and robotic explorers [Stubbs et al., 2007a; 
Farrell et al., 2008].

\vskip 20pt
\noindent{\tenbf 3. The Effective Dust Cross Section}

\nobreak

\vskip 5pt

\nobreak
Although neutral solar wind also forms through interaction with neutral atoms,
we will calculate here only the flux of neutral solar wind created from
the penetration of exospheric dust. Further, we assume that lunar dust grains 
are spherical with radius $a$, although lunar dust can actually be quite irregular [Carrier et al., 1991; Liu et al., 2007]. However, there has been no direct characterization of exospheric dust. We note that in the following 
calculations the irregular shape of lunar dust grains will tend to increase the flux of neutral solar wind because a larger fraction of the dust grains of a given mass will be penetrable.

We consider solar wind protons, which range in speed from about 200~km/s (209~eV) to 1200~km/s (7.5~keV) with a nominal value of about 437~km/s
(1~keV). If a solar wind proton impacting a dust grain has a range $\lambda$ (typically 0.01~$\mu$m to 0.1~$\mu$m) in the dust grain, then provided
$\lambda > 2a$ the proton can penetrate the dust grain regardless of where it
hits (i.e. independent of its impact parameter). Thus, for $\lambda > 2a$, the effective dust cross section is
$\pi a^2$, the projected area of the grain.

However, if $\lambda < 2a$, then the proton can only penetrate the fraction of the dust grain's projected area around the edge of the grain where its path length does not exceed its range. If the proton's path through the grain has a minimum distance to the grain's center (i.e. impact parameter) of $r$, then it must penetrate a length of $2\sqrt{a^2-r^2}$ to make it through the dust grain,
such that the requirement for dust penetration is
$$\eqalignno{ \lambda &> 2\sqrt{a^2-r^2},\ \ \ \ {\rm or} &(1)\cr
r &> \sqrt{a^2-{({\lambda/2})}^2}. &(2)}$$

This means that only the cross-sectional area between $r = \sqrt{a^2-({\lambda/2})}$ and $r = a$ contributes to the effective cross-section such that the effective cross-section, $\sigma_{eff}(a)$ becomes:
$$\eqalignno{ \sigma_{eff}(a) &= \pi a^2 - 
\pi {\biggl[\sqrt{a^2-{({\lambda/2})}^2}\biggr]}^2, \cr
&= \pi {\bigl({\lambda\over 2}\bigr)}^2. &(3)}$$

Thus, 
$$\eqalignno{ \sigma_{eff}(a) &= {\pi\over 4} \lambda^2
\ \ \ \ \ \ \ \ \ \ {\rm for}\ \ \ \lambda < 2a\cr
\sigma_{eff}(a) &= \pi a^2
\ \ \ \ \ \ \ \ \ \ \ {\rm for}\ \ \ \lambda > 2a. &(4)}$$
Interestingly, for larger dust grains ($\lambda < 2a$), 
the cross section is independent of grain radius. 
This is because although larger grains have
a greater total cross-section, a smaller fraction of that total cross-section can be penetrated such that the two effects cancel. 

\vskip 20pt
\noindent{\tenbf 4. The Lunar Dust Concentration at the Terminator}

\nobreak

\vskip 5pt

\nobreak
Murphy and Vondrak [1993] used dust radii and altitudes (scale heights) from Rennilson and Criswell [1974], McCoy [1976], and Zook and McCoy [1991] together
with the assumption that the scale height is a power law function of grain radius, to estimate that
$$\eqalignno{ 
z_0 
&= { {\cal \char'114}\over{a^{8/3}} }, 
&(5)}$$
where $z_0$ is the scale height and $a$ is the dust radius, both in meters, and
${\cal \char'114} = 2{\rm x}10^{-15}\,\, {\rm m}^{11/3}$.

Following the work of Murphy and Vondrak [1993], we assume the dust concentration for a given grain radius above the lunar surface is a power law with an index of
unity, and that exospheric dust has an exponential distribution with height above the surface
($z$) at a given dust radius. However, the estimates of lunar dust distributions from scattered light observations are still controversial, and should be considered ``order of magnitude" as should the predictions presented here. However, note that all our assumptions in the paper tend to be conservative such that they are more likely to underestimate the actual flux of lunar dust-generated neutral solar wind.

Thus, we take the dust concentration per unit dust radius at the terminator to be
$$\eqalignno{ 
\rho(a,z) 
&= {{n_0}\over a} \exp\bigl\{ -{ {a^{8/3} z}\over{\cal \char'114} } \bigr\}, 
&(6)}$$
where altitude $z$ is in meters and $n_0$ is a scaling factor with dimension ${\rm m}^{-3}$.

Using expression~(6), we can calculate the dust column concentration at the terminator
$\Gamma$, for grain radii greater than some $a_{\rm min}$ as
$$\eqalignno{ 
\Gamma &= \int_{a_{\rm min}}^{\infty} \int_0^\infty 
{{n_0}\over a} \exp\bigl\{ -{ {a^{8/3} z}\over{\cal \char'114} } \bigr\} 
\,\, dz\, da\cr
&= {{3\, n_0\, {\cal\char'114}}\over{8\, a_{\rm min}^{8/3}} }. 
&(7)}$$

Murphy and Vondrak [1993] determined $\Gamma$ and $n_0$ using the sky brightness observed during lunar ``twilight" by the astrophotometer aboard Lunokhod-II [Severny et al., 1974], together with the estimates of exospheric dust concentrations from McCoy's [1976] model ``0". The best fit was found for $\Gamma = 1.4{\rm x}10^9~{\rm m}^{-2}$ for dust radii from 0.1~$\mu$m 
to 6~$\mu$m, as reported by Stubbs et al. [2007d].


Equating this value of $\Gamma$ to expression~(7), using $10^{-7}~{\rm m}$ 
(0.1~$\mu$m) as $a_{\rm min}$, and noting that for a maximum dust radius of
$6{\rm x}10^{-6}~{\rm m}$ (6~$\mu$m) the integration to infinity is a good
approximation, we get
$$\eqalignno{ 
n_0 &= 4.0{\rm x}10^5\ \ {\rm m}^{-3}. &(8)}$$

Thus, the expression we will use for the exospheric dust distribution is a modified version of the Murphy and Vondrak [1993] model, 
$$\eqalignno{ 
\rho(a,z) &= {{4{\rm x}10^5}\over a} \exp\bigl\{ -{ {a^{8/3} z}\over{2{\rm x}10^{-15}} } \bigr\},
&(9)}$$
where $\rho$ is in units of ${\rm m}^{-4}$ and $a$ and $z$ are in units of ${\rm m}$.

It is interesting to use expression~(9) to calculate the concentration versus
altitude profile ${\rm n}(z)$ for dust grains larger than some radius
$a_{\rm min}$ at the terminator:
$$\eqalignno{ {\rm n}(z) &= \int_{a_{\rm min}}^\infty \rho(a,z) \, da \cr &=
{4{\rm x}10^5} \int_{a_{\rm min}}^\infty
{{1}\over a} \exp\bigl\{ -{ {a^{8/3} z}\over{2{\rm x}10^{-15}} } \bigr\}\, da, &(10)}$$
where $n$ is in units of ${\rm m}^{-3}$.
By using the change of variable 
$$\eqalignno{ x &= { {a^{8/3} z}\over{2{\rm x}10^{-15}} },
&(11)}$$
this integral may be recast as
$$\eqalignno{ {\rm n}(z) &= {3\over 8} \cdot {4{\rm x}10^5}
\int_{{a_{\rm min}^{8/3} z}\over{2{\rm x}10^{-15}}}^\infty
{1\over x} \exp\bigl\{ {-x} \bigr\}\, dx
\cr &=
{1.5{\rm x}10^5}\,\, {\rm E}_1\bigl(
{{a_{\rm min}^{8/3} z}\over{2{\rm x}10^{-15}}} \bigr),
&(12)}$$
where 
$$\eqalignno{ {\rm E}_1\bigl(x\bigr) &= \Gamma(0,x) \cr &=
\int_x^\infty \exp\{ {-t} \} t^{-1}\, dt \cr &=
-\gamma - \ln\{x\} - \sum_{n=1}^{\infty} { {(-1)^n x^n}\over
{n n!} }, &(13)}$$
where $\gamma = 0.577$ is Euler's constant [Abramowitz and Stegun, 1972].

Thus, for $a_{\rm min} = 0.1{\rm x}10{}^{-6}$~m, we get
$$\eqalignno{ {\rm n}(z) &= 
{1.5{\rm x}10^5}\,\, {\rm E}_1\bigl(
1.1{\rm x}10^{-4} \cdot z\bigr).
&(14)}$$
Therefore, as long as $z {\ll} (1.1{\rm x}10^{-4}~{\rm m}^{-1})^{-1} = 9.3~{\rm km}$, we can use the approximation
$$\eqalignno{ {\rm E}_1\bigl(x\bigr) &\approx -\gamma-\ln\{x\}.
&(15)}$$
Thus,
$$\eqalignno{ {\rm n}(z) &= 
{1.5{\rm x}10^5}\,\, \bigl[
{-\gamma - \ln\{1.1{\rm x}10^{-4} z\} }\bigr] \cr
&= {1.5{\rm x}10^5}\,\, \bigl[
{-\ln\{e^\gamma\} - \ln\{1.1{\rm x}10^{-4} z\} }\bigr] \cr
&= {1.5{\rm x}10^5}\,\,
\ln\bigl\{{{5100}\over z}\bigr\}.
&(16)}$$

Figure~1 shows the exospheric dust concentration as a function of altitude at the lunar terminator from Stubbs et al. [2007d] along with the approximation given by equation~(16). The approximation appears to be reasonable within a factor 
of two or so, at least up to 1~km altitude.

\vskip 20pt
\noindent{\tenbf 5. ``Visible" Light-Scattering Dust Model}

\nobreak

\vskip 5pt

\nobreak
In this model, we assume that the smallest lunar dust grains are those that 
are effective at scattering visible light ($a_{min} = 0.1~\mu{\rm m}$).
We will also assume the composition of lunar grains is silicon
mainly because data on particle range through silicon are readily available. However, this seems a reasonable approximation as even mafic materials have a very high silica content. The composition of lunar dust does change with dust radius: crystalline silica and plagioclase/mafic mineral ratios are greater in the $a < 5~\mu{\rm m}$ fraction than in the coarser fractions for all lunar soils. Comminution (breaking into fragments), agglutination (melting), vertical mixing, and local lateral transport control the composition versus radius relationship
[Heiken et al., 1991].

Figure~2, based on the data in figure~4a of Demond et al. [1980], shows the 
range in silicon in $\mu{\rm m}$ as a function of hydrogen speed in km/s. It is not anticipated that there is any substantial difference in the data between hydrogen and proton penetration. The figure suggests that a good approximation for the range of protons in $\mu$m through lunar 
dust is
$$\eqalignno{ \lambda &= 
{8.9{\rm x}10^{-8}}\cdot {v}^2 \cr
&\equiv 0.017~\mu{\rm m}/{\rm keV}, &(17)}$$
where $v$ is the proton speed in km/s.

It should be noted that equation~(17) is an average projected range and that the actual range for any given proton can vary considerably. We use this expression as a reasonable estimate with some protons penetrating farther and some not as far. Also, this will depend on the shape characteristics of exospheric dust which are not well-known.
This means that, considering only ``visible" light-scattering dust, all reasonable solar wind velocities fall in the regime for $\lambda<2a$ (see equation~4) such that
$$\eqalignno{ \sigma_{eff}(a) &= {\pi\over 4} \lambda^2\cr
&= {6.2{\rm x}10^{-27}} {v}^4, &(18)}$$
where $\sigma_{eff}$ is in m${}^2$ and $v$ is in km/s.

Based on the data from Figure~1 of Kallenbach et al. [1993], 
at these energies, $>$80\%
of the protons exiting thin carbon foils are neutral. Although it is not
clear whether these data are directly applicable to this
problem, the charge state distribution of protons through lunar dust is
likely similar [Wimmer-Schweingruber and Bochsler, 2003]. We will assume because such a large fraction of protons exiting carbon foils are neutral that all protons that penetrate lunar dust grains emerge neutral.

Figure~3 shows the viewing geometry, which is tangential to the lunar surface at the terminator. Here $x$ is the distance along the line-of-sight viewing into the solar wind at the terminator, $R$ is the lunar radius, 1738~km, and $z$ is the height above the lunar surface of the point $x$. Thus,
$$\eqalignno{ (z+R)^2 &= x^2 + R^2, &(19)}$$
or
$$\eqalignno{ z &= R\sqrt{1 + x^2/R^2} - R \cr
&\approx R\biggl[1 + {x^2\over{2 R^2}} \biggr] - R \cr
&= {x^2\over{2 R}}, &(20)}$$
where we have expanded to first order under the assumption that $x\ll R$. This expansion is not formally correct for very small grains which have a large scale height. However, equation~(20) is still sufficient for our purposes (see Section~6). 

We can calculate the ratio of the neutral solar wind flux $\Phi_{\rm NSW}$, to
solar wind flux $\Phi_{\rm sw}$, by integrating over all dust radii along the line-of-sight [Roelof, 1997; Roelof and Skinner, 2000]:
$$\eqalignno{ {\Phi_{\rm NSW}}\over{\Phi_{\rm sw}} &= 
\int_{a_{\rm min}}^{\infty} \int_0^\infty   
{{n_0}\over a} \exp\bigl\{ -{ {a^{8/3} z}\over{\cal \char'114} } \bigr\} 
\cdot \sigma_{eff}(a)\,\, dx\, da. &(21)}$$
In expression~(21), we have used the dust distribution function given by equation~(9) and taken the effective cross-section to depend only on the
proton range because of the relatively large radius of the dust grains considered here:
$$\eqalignno{ \sigma_{eff}(a) &= 
{\pi\over 4} \lambda^2. &(22)}$$
Thus,
$$\eqalignno{ {\Phi_{\rm NSW}}\over{\Phi_{\rm sw}} &= {\pi\over 4} \lambda^2 
\int_{a_{\rm min}}^{\infty} {{n_0}\over a}
\int_0^\infty \exp\bigl\{ -{ {a^{8/3}\over{\cal \char'114}} \cdot
{x^2\over{2 R}} } \bigr\} \,\, dx\, da. &(23)}$$
Noting that integrating the normalized Gaussian gives
$$\eqalignno{ \int_{-\infty}^{\infty} \exp\bigl\{ -{x^2\over{2\sigma^2}} \bigr\}
\,\, dx &= \sqrt{2\pi}\, \sigma, &(24)}$$
the {\it x}-integral above becomes
$$\eqalignno{ \int_0^\infty \exp\bigl\{ -{ {a^{8/3}\over{\cal \char'114}} \cdot
{x^2\over{2 R}} } \bigr\} \,\, dx &= 
\sqrt{ {\pi {\cal \char'114} R}\over 2}\, a^{-4/3}. &(25)}$$
Consequently,
$$\eqalignno{ {\Phi_{\rm NSW}}\over{\Phi_{\rm sw}} &= 
{ {\pi^{3/2}\, \lambda^2\, n_0}\over 4 } \sqrt{ {{\cal \char'114} R}\over 2}
\int_{a_{\rm min}}^{\infty} a^{-7/3} \,\, da \cr
&= { 3{\pi^{3/2}\, \lambda^2\, n_0}\over 16 } \sqrt{ {{\cal \char'114} R}\over 2}
\cdot {1\over a_{\rm min}^{4/3}}. 
&(26)}$$
Using the values $\lambda = 8.9{\rm x}10^{-14}\, v^2\, {\rm m}$ from equation~(17), $n_0 = 4.0{\rm x}10^5\, {\rm m}^{-3}$ from equation~(8), ${\cal \char'114} = 2.0{\rm x}10^{-15}\, {\rm m}^{11/3}$, $R = 1738{\rm x}10^3\, {\rm m}$, and $a_{\rm min} = 0.1{\rm x}10^{-6}\, {\rm m}$ for the smallest dust grain radius considered here (which the solar wind is taken to not fully penetrate), we get
$$\eqalignno{ {\Phi_{\rm NSW}}\over{\Phi_{\rm sw}} &=
3.0{\rm x}10^{-16}\, v^4. &(27)}$$

The solid line in Figure~4 shows this ratio plotted as a function of solar wind speed. At high solar wind speeds, even considering only ``large" dust grains, the ratio of neutral solar wind to solar wind flux comes close to 10${}^{-3}$. Note that in this expression, because the solar wind does not completely penetrate the grains for any reasonable speed, we never reach a regime in which the neutral solar wind flux is directly proportional to solar wind flux. Interestingly, this non-linear behavior also exists for magnetosheath charge exchange at the Earth [Collier et al., 2005] and may very well be a general property of neutral solar wind production. 

For reference, the dotted line in Figure~4 shows the neutral solar wind count rate that an instrument like IMAGE/LENA [Moore et al., 2000] would observe at the terminator assuming a hydrogen efficiency of $2{\rm x}10^{-4}$, an aperture size of 1~cm${}^2$, and an average solar wind flux of $3{\rm x}10^8$/cm${}^2$/s.

\vskip 20pt
\noindent{\tenbf 6. ``Ultraviolet" Scattering Dust Model}

\nobreak

\vskip 5pt

\nobreak
In this model, we assume that the smallest lunar dust grains are those that are
more effective at scattering ultraviolet light ($a_{min} = 0.01~\mu$m).
It is believed that
grains of this radius are easily lost since they are prone to tribocharging and attach to everything. Hence, when the Apollo samples were transferred between containers, it is highly likely that
the smallest grains were preferentially lost, which would result in their 
being under-represented in any subsequent size distribution analysis. 
However, 
UV horizon glow seen by Page and Carruthers [1978] in the 1050\AA\  to 1600\AA\        
(105 to 160 nm) range could be interpreted as
due to dust as small as 0.01~$\mu$m. Furthermore, grains this small are present in the Apollo samples [Park et al., 2006].

In order to fully penetrate the smallest grains in the population, 
$\lambda > 2a$:
$$\eqalignno{ 2a\,\, &=\,\, 0.02\,\mu{\rm m}\,\, <\,\, 
8.9{\rm x}10^{-8} {v}^2, &(28)}$$
or
$$\eqalignno{ {v}\,\, &>\,\, 474~{\rm km/s}\,\, =\,\, v_{crit}, &(29)}$$
so that below about average solar wind speeds, the solar wind does not fully penetrate any of the lunar dust grains, and we can use the formalism for the ``visible" dust model, but of course with the neutral solar wind fluxes scaled upwards by a factor of $10^{4/3} = 21.5$ because the minimum grain radius in the ``UV" model is a factor of 10 smaller than the ``visible" model (0.01~$\mu$m versus 0.1~$\mu$m) described by equation~(26). Thus, 
$$\eqalignno{ {\Phi_{\rm NSW}}\over{\Phi_{\rm sw}} &=
6.4{\rm x}10^{-15}\, v^4, \,\,\,\,\,\,\,\,\,\, 
v<474~{\rm km/s} &(30)}$$

For solar wind velocities in excess of $v_{crit}$, the solar wind fully penetrates lunar dust grains with radii smaller than $a_{tr}$, so-called because it marks the ``transition" radius from partial to full dust penetration, where
$$\eqalignno{ a_{tr} \,\, &= 
4.5{\rm x}10^{-8} {v}^2, &(31)}$$
from equation~(28) with {\it v} again in km/s and $a_{tr}$ in $\mu$m.
For dust grain radii smaller than $a_{tr}$, the solar wind fully penetrates the grains and the effective cross section
$$\eqalignno{ \sigma_{eff}(a) &= \pi a^2, \,\,\,\,\,\,\,\,\,\,
a \le a_{tr} &(32)}$$
while for grain radii in excess of $a_{tr}$, the effective cross section is
$$\eqalignno{ \sigma_{eff}(a) &= {\pi\over 4} \lambda^2 \cr
&= {\pi\over 4}\cdot 7.9{\rm x}10^{-27} {v}^4,
\,\,\,\,\,\,\,\,\,\, a>a_{tr} &(33)}$$
with {\it v} in km/s. This is, of course, just $\pi a_{tr}^2$ because the cross section is independent of grain radius. 

So, for the flux ratio at solar wind speeds in excess of $v_{crit}$, we simply start with equation~(21),
$$\eqalignno{ {\Phi_{\rm NSW}}\over{\Phi_{\rm sw}} &= 
\int_{a_{\rm min}}^{\infty} \int_0^\infty   
{{n_0}\over a} \exp\bigl\{ -{ {a^{8/3} z}\over{\cal \char'114} } \bigr\} 
\cdot \sigma_{eff}(a)\,\, dx\, da\cr
&= \int_{a_{\rm min}}^{\infty} {{n_0}\over a}\,\, \sigma_{eff}(a)\,\,
\int_0^\infty \exp\bigl\{ -{ {a^{8/3}\over{\cal \char'114}} \cdot
{x^2\over{2 R}} } \bigr\} \,\, dx\, da\cr
&= n_0\,\, \sqrt{{\pi {\cal \char'114} R}\over 2}
\int_{a_{\rm min}}^{\infty} \sigma_{eff}(a) \cdot a^{-7/3} \,\, da, &(34)}$$
but note that the cross section, $\sigma_{eff}(a)$,
now depends on grain radius with the dependence beginning at grain radii smaller
than $a_{tr}$:
$$\eqalignno{ {\Phi_{\rm NSW}}\over{\Phi_{\rm sw}} &= 
n_0\,\, \sqrt{{\pi {\cal \char'114} R}\over 2} \cdot
\biggl[ \int_{a_{\rm min}}^{a_{\rm tr}} \pi a^2 \cdot a^{-7/3}\,\, da +
\int_{a_{\rm tr}}^{\infty} {\pi\over 4} \lambda^2 \cdot a^{-7/3}\,\, da
\biggr]\cr
&= n_0 \pi^{3/2} \sqrt{{{\cal \char'114} R}\over 2} \biggl[
\int_{a_{\rm min}}^{a_{\rm tr}} a^{-1/3}\,\, da +
{{\lambda^2}\over 4} 
\int_{a_{\rm tr}}^{\infty} a^{-7/3}\,\, da \biggr] \cr
&= n_0 \pi^{3/2} \sqrt{{{\cal \char'114} R}\over 2} \cdot {3\over 2}
\bigl[ a^{2/3}\big\vert_{a_{\rm min}}^{a_{\rm tr}} \bigr] +
{{\pi^{3/2} \lambda^2 n_0}\over 4} \sqrt{{{\cal \char'114} R}\over 2} \cdot
\bigl(-{3\over 4}\bigr) \bigl[ a^{-4/3}\big\vert_{a_{\rm tr}}^{\infty} \bigr] \cr
&= {{3\pi^{3/2}\, n_0}\over 2} \sqrt{{{\cal \char'114} R}\over 2}
\biggl[ a_{\rm tr}^{2/3} - a_{\rm min}^{2/3} + {\lambda^2\over 8}
\cdot {1\over a_{\rm tr}^{4/3}} \biggr]
&(35)}$$
This expression can be re-written as
$$\eqalignno{ {\Phi_{\rm NSW}}\over{\Phi_{\rm sw}} &= 
{{3\pi^{3/2}\, n_0}\over 2} \sqrt{{{\cal \char'114} R}\over 2}
{{\,\,\lambda^2}\over 8} \cdot {1\over{a_{\rm min}^{4/3}}} \cdot
\biggl[ {8\over {\lambda^2}} \bigl[ 
a_{\rm min}^{4/3} a_{\rm tr}^{2/3} - a_{\rm min}^{2} \bigr] +
{\bigl( {a_{\rm min}\over a_{\rm tr}} \bigr)}^{4/3} \biggr],
\,\,\,\,\,\,\,\,\,\, v>v_{crit} &(36)}$$
where the expression outside of the large square brackets to the right of the equals sign is expression~(26) and represents the flux ratio achieved if the protons do not fully penetrate the grains. The expression inside the large square brackets, then, may be viewed as a ``correction term" which takes into account that the cross section does not increase with increasing solar wind speed once the solar wind fully penetrates a particular grain size. Noting that $\lambda = 2 a_{\rm tr}$, the term in the square brackets $C(a_{\rm tr})$, may be expressed as
$$\eqalignno{ C(a_{\rm tr}) &= 
3 {\biggl( {a_{\rm min}\over a_{\rm tr}} \biggr)}^{4/3} -
2 {\biggl( {a_{\rm min}\over a_{\rm tr}} \biggr)}^{2},
&(37)}$$
which, because $a_{\rm tr}\ge a_{\rm min}$, $C(a_{\rm tr})\le 1$.
Figure~5 shows this expression as a function of 
${a_{\rm tr}/a_{\rm min}}$. Taking $a_{\rm min}=0.01{\rm x}10^{-6}$~m and
$a_{\rm tr} = 4.5{\rm x}10^{-14} {v}^2$~m with v in km/s, we get
$$\eqalignno{ C(v) &= 
{{4.0{\rm x}10^7}\over{{v}^{8/3}}} -
{{9.7{\rm x}10^{10}}\over{{v}^{4}}}. &(38)}$$
Using the values from above, namely, $\lambda = 8.9{\rm x}10^{-14}\, v^2\, {\rm m}$ from equation~(17), $n_0 = 4.0{\rm x}10^5\, {\rm m}^{-3}$ from equation~(8), ${\cal \char'114} = 2.0{\rm x}10^{-15}\, {\rm m}^{11/3}$, and $R = 1738{\rm x}10^3\, {\rm m}$, but now taking $a_{\rm min} = 0.01{\rm x}10^{-6}\, {\rm m}$ for the smallest dust grain radius considered here, we get
$$\eqalignno{ {\Phi_{\rm NSW}}\over{\Phi_{\rm sw}} &=
6.4{\rm x}10^{-15}\, v^4
\cdot C(v), \,\,\,\,\,\,\,\,\,\, 
v>474~{\rm km/s} &(39)}$$
the difference between this and equation~(30) being the presence of the factor
$C(v)$. Thus, we have for the UV model:
$$\eqalignno{ 
{\Phi_{\rm NSW}}\over{\Phi_{\rm sw}} &=
6.4{\rm x}10^{-15}\, v^4, 
\,\,\,\,\,\,\,\,\,\,\,\,\,\,\,\,\,\,\,\,\,\,\,\,\,\,\,\,\,\,\,\,\,\,\,\,\,\,\,\, 
v<474~{\rm km/s} = v_{crit} \cr
{\Phi_{\rm NSW}}\over{\Phi_{\rm sw}} &=
6.4{\rm x}10^{-15}\, v^4
\cdot C(v), \,\,\,\,\,\,\,\,\,\,\,\,\,\,\,\,\,\,\,\,\,\,\,
v>474~{\rm km/s} = v_{crit} &(40)}$$

Figure~6 shows the ratio of neutral solar wind flux to solar wind flux from equation~(40) on the left y-axis and solid line assuming the minimum dust grain 
radius is 0.01$\mu{\rm m}$. The right y-axis and dotted line shows the corresponding count rate assuming the same parameters as in Figure~4. At the highest solar wind speeds, a LENA-like instrument would observe a count rate of over 100~$s^{-1}$. For all except the very slowest solar wind speeds, 
such an instrument would register a count rate $> 1\,\,\, s^{-1}$. 

Note that, although the solar wind is typically about 96\% protons by number, smaller
neutral atom fluxes of minor species, notably helium and oxygen, would also be
created by the solar wind interaction with lunar dust.

The validity of the approximation $z=x^2/2R$ used in equation~(20) is dependent on the maximum dust scale height in comparison to the lunar radius. This scale height, ${\cal \char'114}/a_{\rm min}^{8/3}$, is only 9.3~km for the ``visible" light-scattering dust model discussed in the previous section corresponding to $a_{min}=10{}^{-7}$~m and so the approximation is valid.
However, for the ``UV" scattering model, this scale height is 4300~km (10${}^{8/3}$ = 464 times larger) which is about 2.5 times larger than the lunar radius of 1738~km. However, the overall accuracy of the approximation will be bounded by how well the spatial integral using $z=x^2/2R$ compares to the integral, evaluated numerically, without this approximation. The spatial integral evaluated numerically as compared with the approximation in equation~(25) given by $7.4{\rm x}10^{-5} a^{-4/3}$ as a function of dust radius shows that the maximum discrepancy occurs at the smallest dust radius, as expected, and is only about 20\%. Worth noting is that the approximation actually underestimates the spatial integral such that the actual flux of lunar dust-generated neutral solar wind will be larger than predicted by the analytic solution.

\vskip 20pt
\noindent{\tenbf 7. Discussion}

\nobreak

\vskip 5pt

\nobreak
It has been the purpose of this paper to examine neutral solar wind created by lunar exospheric dust. However, as mentioned in section~1, neutral solar wind also results from solar wind charge exchange (SWCX) with exospheres [e.g. Collier et al., 2005] and this process also occurs at the Moon. 
If we assume an argon atmosphere with a surface density of $n_{Ar} = 10^5$~cm${}^{-3}$, a scale height of $H_{Ar} = 40$~km and a canonical collision cross section $\sigma_{Ar} = 
10^{-15}$~cm${}^2$ [Stern, 1999], one can estimate a ratio of neutral solar wind to solar wind flux viewing normal to the surface $\Phi_{\rm NSW}/\Phi_{\rm sw} \sim n_{Ar} \sigma_{Ar} H_{Ar} = 4 {\rm x} 10^{-4}$. 
Assuming limb viewing, $\Phi_{\rm NSW}/\Phi_{\rm sw}$ will be $\sim10^{-3}$.
Referring to Figure~6, at higher solar wind speeds the lunar dust-generated neutral solar wind flux will likely exceed that due to charge exchange while at slower solar wind speeds the opposite will be the case.

However, even at slower solar wind speeds there are a number of factors which would allow a separation between charge-exchange and lunar dust-generated neutral solar wind: (i) Dust-generated  energetic neutral atoms (ENAs) would not be as beam-like as the charge-exchange generated ENAs because penetrating dust grains will scatter the resulting neutrals in angle; (ii) Lunar dust-generated neutral solar wind would contain species other than just hydrogen because penetrating dust can neutralize high charge state ions; (iii) Lunar dust-generated neutral solar wind would have a significantly wider energy distribution than charge-exchange generated neutral solar wind because the process of penetrating dust grains involves energy loss and straggling; (iv) Finally, the relative flux of dust-generated neutral solar wind increases dramatically with solar wind speed compared with charge exchange with neutral exospheric atoms due to the solar wind speed-dependent cross-section [e.g. Rucinski et al., 1996].

Along similar lines, the ratio of the neutral solar wind to ionized solar wind due to charge transfer in the interplanetary medium (i.e. without an exosphere) is expected to be $10^{-5}-10^{-4}$ [e.g. Holzer, 1977; Gruntman, 1994], so that the lunar dust-generated neutral solar wind signal is comparable to this population at slow solar wind speeds but dominates it at higher solar wind speeds. Furthermore, because the lunar dust-generated neutral solar wind is 
anticipated to be a very local phenomenon, changes in its flux will be more highly correlated than the ``background" neutral solar wind with ionized solar wind observations from an upstream monitor within a few tens of Earth radii from the location of the Moon [e.g. Collier et al., 1998; Richardson and Paularena, 2001].

There were a number of simplifying assumptions that went into the analytic results in the previous sections. Among these is the use of data on the penetration of protons through silicon which were readily available 
due to the commercial importance of implanted silicon. However, a more accurate approach might be to use data for the penetration of protons through silica [Eder et al., 1997]. Although the data are not as complete as those for protons through silicon, the data suggest the range is higher for protons through silica. We also assumed that lunar dust is spherical when in reality its shape is more complex. These assumptions may lead to a neutral solar wind flux underestimated by a factor of four or more.

Some of the other assumptions are that protons execute straight-line trajectories through a dust grain. Although this is a reasonable first approximation, some protons will back-scatter from the dust grain while those that penetrate 
will both scatter in angle and lose energy. Furthermore, the penetrating protons will sputter both ions and neutral atoms from the dust grains. 
None of these effects are well-understood. For example, the sputtering yield formulae may not be accurate for small spherical dust grains because they are obtained from experiments on flat solids. Furthermore, transmission sputtering (sputtering on exit) may be important, but has not been investigated in the regime of interest here [e.g. Dwek and Arendt, 1992]. However, the sputtered products will have relatively low energies ($\sim$10~eV) in comparison to the solar wind [Johnson and Sittler, 1990; Elphic et al., 1991]. Consequently, they should be easily separated out from the more energetic neutral solar wind component. 

The energy distribution will also be affected by the solar wind-dust interaction. Ions traversing through matter lose energy by a sequence of
stochastic processes which result in an energy loss and dispersion. Typically at 
incident energies around 4 keV, proton beams traversing solid foils between 0.01 and 0.03 $\mu$m thickness will emerge with $\sim$3~keV with energy dispersions of the order of a few hundred eV [Figueroa et al., 2007]. So, the observed energy distribution of the lunar dust-generated neutral solar wind will have lower energies than that of the solar wind and will be substantially hotter.
Also note that energy arguments suggest that a 1~keV proton depositing all its energy into a 0.01~$\mu$m dust grain will not destroy the dust grain or significantly affect the overall integrity of the grain.

\vskip 20pt
\noindent{\tenbf 8. Conclusions}

\nobreak

\vskip 5pt

\nobreak
Dust is a recognized planetary hazard which will have an adverse impact on robotic and human exploration of the Moon and Mars. In this paper, we showed that based on our current understanding of lunar exospheric dust characteristics at the terminator, a significant and easily observable flux of energetic neutral solar wind hydrogen, at times higher than 10${}^{-3}$ the solar wind flux even
when conservatively estimated, will
be present in the terminator region. Thus, neutral atom observations would provide information on the exospheric dust distribution at the Moon. For example, the distribution of small grains, which are easier for the solar wind to penetrate, could be determined if one assumes the altitude profile follows the Murphy and Vondrak [1993] distribution used in the paper. Furthermore, other model distributions could be tested [e.g. Stubbs et al., 2007d].
Consequently, neutral atom observations would be highly complementary to optical techniques for characterizing lunar dust and should be considered an important component of any lunar robotic mission designed to assess the characteristics of this hazard.

It has been speculated that activities associated with in-situ resource 
utilization (ISRU) could increase the amount of exospheric lunar dust by several orders of magnitude [Vondrak, 2007]. Given that these activities may very well occur at the polar regions where the terminator is omnipresent and that the 
ratio of neutral solar wind flux to solar wind flux increases linearly with dust 
concentration (i.e. $n_0$), such activities may very well neutralize a large fraction of the solar wind.

Retarded space weathering due to magnetic anomalies shielding the lunar
surface from the solar wind has been proposed as an explanation for 
high albedo markings [e.g. Hood and Williams, 1989]. Because neutral solar
wind will not be deflected by magnetic anomalies, a substantial flux of
neutral solar wind could alter space weathering processes.

\vskip 20pt
\noindent
{\tenbf Acknowledgments.} We thank both of the referees for very helpful comments. Thanks also to Pamela Clark, Dave Glenar, Bill Farrell, John Keller, and Richard Vondrak for helpful conversations. We gratefully acknowledge the support of NASA/GSFC IRAD funding. 
\vfill \eject

\vskip 20pt
\noindent {\tenbf References}

\nobreak

\vskip 5pt

\nobreak

\hangindent=1.5pc \hangafter=1 \noindent
{Abramowitz, M. and I. Stegun (1972), Handbook of Mathematical Functions with Formulas, Graphs, and Mathematical Tables, Dover Publications, New York, 228.}

\hangindent=1.5pc \hangafter=1 \noindent
{Berg, O.E., H. Wolf and J. Rhee. (1976), Lunar soil movement registered by the Apollo~17 cosmic dust experiment, in {\it Interplanetary Dust and Zodiacal Light}, Proceedings of the Colloqium, 31, Springer-Verlag, Berlin, 233-237.}

\hangindent=1.5pc \hangafter=1 \noindent
{Borisov, N. and U. Mall (2006), Charging and motion of dust grains near the terminator of the moon, {\tenit Planet. Space Sci., 54\/}, 572-580, doi:10.1016/j.pss.2006.01.005.}

\hangindent=1.5pc \hangafter=1 \noindent
{Brinkfeldt, K. and 46 coauthors (2006), First ENA observations at Mars: Solar-wind ENAs on the nightside, {\tenit Icarus, 182\/}, 439-447.}

\hangindent=1.5pc \hangafter=1 \noindent
{Carrier, W. David III, G.R. Olhoeft and W. Mendell (1991), Physical properties of the lunar surface, in {\it Lunar Sourcebook}, G.H. Heiken, D.T. Vaniman and B.M. French (eds.), 478.}

\hangindent=1.5pc \hangafter=1 \noindent
{Collier, Michael R., J.A. Slavin, R.P. Lepping, A. Szabo and K. Ogilvie (1998), Timing accuracy for the simple planar propagation of magnetic field structures in the solar wind, {\tenit Geophys. Res. Lett., 25\/}, 2509-2512.}

\hangindent=1.5pc \hangafter=1 \noindent
{Collier, Michael R., Thomas E. Moore, Keith W. Ogilvie, Dennis Chornay, J.W. Keller, S. Boardsen, James Burch, B. El Marji, M.-C. Fok, S.A. Fuselier, A.G. Ghielmetti, B.L. Giles, D.C. Hamilton, B.L. Peko, J.M. Quinn, E.C. Roelof, T.M. Stephen, G.R. Wilson, and P. Wurz (2001), Observations of neutral atoms from the solar wind, {\tenit J. Geophys. Res., 106\/}, 24,893-24,906.}

\hangindent=1.5pc \hangafter=1 \noindent
{Collier, Michael R., Thomas E. Moore, K. Ogilvie, D.J. Chornay, J. Keller, S. Fuselier, J. Quinn, P. Wurz, M. Wuest and K.C. Hsieh (2003), The dust geometric cross section at 1~AU based on neutral solar wind observations, {\tenit Solar Wind Ten\/}, Proceedings of the Tenth International Solar Wind Conference, M. Velli, R. Bruno and F. Malara (eds.), AIP Conf. Proceedings, Volume 679, American Institute of Physics, Melville, New York, pp. 790-793.}

\hangindent=1.5pc \hangafter=1 \noindent
{Collier, M. R., T. E. Moore, M.-C. Fok, B. Pilkerton, S. Boardsen, and
H. Khan, (2005), Low energy neutral atom signatures of magnetopause
motion in response to southward B${}_z$, {\tenit J. Geophys. Res., 110\/}, 
A02102, doi:10.1029/2004JA010626.}

\hangindent=1.5pc \hangafter=1 \noindent
{Colwell, J.E., S. Batiste, M. Horanyi, S. Robertson and S. Sture (2007), Lunar surface: Dust dynamics and regolith mechanics, {\tenit Rev. of Geophys., 45\/}, RG2006.}

\hangindent=1.5pc \hangafter=1 \noindent
{Criswell, D.R. and B.R. De (1977), Intense localized charging in the lunar sunset terminator region: Supercharging at the progression of sunset, {\tenit J. Geophys. Res., 82\/}, 1005.}

\hangindent=1.5pc \hangafter=1 \noindent
{De, B.R. and D.R. Criswell (1977), Intense localized photoelectric charging in the lunar sunset terminator region\ \ \ \ 1. Development of potentials and fields, {\tenit J. Geophys. Res., 82\/}, 999-1004.}

\hangindent=1.5pc \hangafter=1 \noindent
{Demond, F.-J., S. Kalbitzer, H. Mannsperger and G. M\"uller (1980), Range parameters of protons in silicon implanted at energies from 0.5 to 300~keV, {\tenit Nucl. Inst. and Methods, 168\/}, 69-74.}

\hangindent=1.5pc \hangafter=1 \noindent
{Dwek, E. and R.G. Arendt (1992) Dust-gas interactions and the infrared emiaaion from hot astrophysical plasmas, {\tenit Annu. Rev. Astron. Astrophys., 30\/}, 11-50.}

\hangindent=1.5pc \hangafter=1 \noindent
{Eder, K., D. Semrad, P. Bauer, R. Golser, P. Maier-Komor, F. Aumayr, M. Pe\~nalba, A. Arnau, J.M. Ugalde and P.M. Echenique (1997), Absence of
a ``threshold effect" in the energy loss of slow protons traversing large-band-gap insulators, {\tenit Phys. Rev. Lett., 79\/}, 4112-4115.}

\hangindent=1.5pc \hangafter=1 \noindent
{Elphic, R.C., H.O. Funsten III, B.L. Barraclough, D.J. McComas, M.T. Paffett, D.T. Vaniman and G. Heiken (1991) Lunar surface composition and solar wind-induced secondary ion mass spectrometry, {\tenit Geophys. Res. Lett., 18(11)\/}, 2165-2168.}

\hangindent=1.5pc \hangafter=1 \noindent
{Fahr, H.J. (1968) Neutral corpuscular energy flux by charge-transfer collisions in the vicinity of the Sun, {\tenit Astrophys. Space Sci., 2\/}, 496-503.}

\hangindent=1.5pc \hangafter=1 \noindent
{Farrell, W.M., T.J. Stubbs, R.R. Vondrak, G.T. Delory, and J.S. Halekas (2007), Complex electric fields near the lunar terminator: The near-surface wake and accelerated dust, {\tenit Geophys. Res. Lett., 34\/}, L14201, doi:10.1029/2007GL029312.}

\hangindent=1.5pc \hangafter=1 \noindent
{Farrell, W.M., T.J. Stubbs, G.T. Delory, R.R. Vondrak, M.R. Collier, J.S. Halekas and R.P. Lin (2008), Concerning the dissipation of electrically-charged objects in the shadowed lunar polar regions, {\tenit Geophys. Res. Lett.\/}, in press.}

\hangindent=1.5pc \hangafter=1 \noindent
{Figueroa, E.A., N.R. Arista, J.C. Eckardt, and G.H. Lantschner (2007), Determination of the difference between the mean and the most probable energy loss of low-energy proton beams traversing thin solid foils, {\tenit Nucl. Inst. Methods B, 256\/}, 126-130.}

\hangindent=1.5pc \hangafter=1 \noindent
{Freeman, J.W. and M. Ibrahim (1975), Lunar electric fields, surface potential and associated plasma sheaths, {\tenit The Moon, 8\/}, 103-114.}

\hangindent=1.5pc \hangafter=1 \noindent
{Gruntman, M.A. (1994), Neutral solar wind properties: Advance warning of major geomagnetic storms, {\tenit J. Geophys. Res., 99\/}, 19,213-19,227.}

\hangindent=1.5pc \hangafter=1 \noindent
{Halekas, J.S., G.T. Delory, D.A. Brain, R.P. Lin, M.O. Fillingim, C.O. Lee, R.A. Mewaldt, T.J. Stubbs, W.M. Farrell and M.K. Hudson (2007), Extreme lunar surface charging during solar energetic particle events, {\tenit Gephys. Res. Lett., 34\/}, L02111, doi:10.1029/2006GL028517.}

\hangindent=1.5pc \hangafter=1 \noindent
{Heiken, G.H., D.T. Vaniman and B.M. French (1991), Lunar Sourcebook: A User's Guide to the Moon, Cambridge University Press and Lunar and Planetary Institute, Cambridge, UK.}

\hangindent=1.5pc \hangafter=1 \noindent
{Holzer, T.E. (1977), Neutral hydrogen in interplanetary space, 
{\tenit Rev. Geophys., 156\/}, 467-490.}

\hangindent=1.5pc \hangafter=1 \noindent
{Hood, L.L. and C.R. Williams (1989), The lunar swirls: Distribution and
possible origins,
{\tenit Proc. of the 19${}^{th}$ Lunar and Planetary Science 
Conference\/}, 99-113.}

\hangindent=1.5pc \hangafter=1 \noindent
{Johnson, R.E. and E.C. Sittler Jr. (1990), Sputter-produced plasma as a measure of satellite surface composition: The Cassini mission, {\tenit Geophys. Res. Lett., 17(10)\/}, 1629-1632.}

\hangindent=1.5pc \hangafter=1 \noindent
{Kallenbach, R., M. Gonin, A. B\"urgi and P. Bochsler (1993), Charge exchange of hydrogen ions in carbon foils, {\tenit Nucl. Inst. and Methods in Phys. Res. B, 83\/}, 68-72.}

\hangindent=1.5pc \hangafter=1 \noindent
{Liu, Y., D. Schnare, J.S. Park, E. Hill, B. Eimer and L.A. Taylor (2007), Shape analysis of lunar dust particles for astronaut toxicological studies, {\tenit Lunar and Planetary Sci. 38${}^{\rm th}$\/}, Abstract 1383.}

\hangindent=1.5pc \hangafter=1 \noindent
{McCoy, J.E. (1976), Photometric studies of light scattering above the lunar terminator from Apollo solar corona photography, {\tenit Proc. Lunar Sci. Conf., 7${}^{\rm th}$\/}, 1087-1112.}

\hangindent=1.5pc \hangafter=1 \noindent
{McCoy, J.E. and D.R. Criswell (1974), {\tenit Proc. Lunar Sci. Conf. 5${}^{\rm th}$\/}, 2991-3005.}

\hangindent=1.5pc \hangafter=1 \noindent
{Moore, T.E. et al. (2000), The Low-Energy Neutral Atom Imager for IMAGE, {\tenit Space Sci. Rev., 91\/}, 155-195.}

\hangindent=1.5pc \hangafter=1 \noindent
{Murphy, D.L., and R.R. Vondrak (1993), Effects of levitated dust on astronomical observations from the lunar surface, {\tenit Proc. Lunar Planet. Sci. Conf. 24${}^{\rm th}$\/}, 1033-1034.}

\hangindent=1.5pc \hangafter=1 \noindent
{Page, T. and G.R. Carruthers (1978), S201 Far-Ultraviolet Atlas of the Large Magellanic Cloud, {\tenit NRL Report 8206\/}, Naval Research Laboratory, Washington, D.C.}

\hangindent=1.5pc \hangafter=1 \noindent
{Park, J.S., Y. Liu, K.D. Kihm and L.A. Taylor (2006), Micro-morphology and lexicological effects of lunar dust, {\tenit Lunar and Planetary Sci. 37${}^{\rm th}$\/}, Abstract 2193.}

\hangindent=1.5pc \hangafter=1 \noindent
{Rennilson, J.J. and D.R. Criswell (1974), Surveyor observations of lunar horizon-glow, {\tenit The Moon, 10\/}, 121-142.}

\hangindent=1.5pc \hangafter=1 \noindent
{Richardson, J.D. and K. Paularena (2001), Plasma and magnetic field correlations in the solar wind, {\tenit J. Geophys. Res., 106\/}, 239-251.}

\hangindent=1.5pc \hangafter=1 \noindent
{Roelof, E.C. (1997), Energetic neutral atom imaging of magnetospheric ions from high and low-altitude spacecraft, {\tenit Adv. Space Res., 20(3)\/}, 341-350.}

\hangindent=1.5pc \hangafter=1 \noindent
{Roelof, E.C. and A.J. Skinner (2000), Extraction of ion distributions from magnetospheric ENA and EUV images, {\tenit Space Sci. Rev., 91\/}, 437-459.}

\hangindent=1.5pc \hangafter=1 \noindent
{Rucinski, D., A.C. Cummings, G. Gloeckler, A.J. Lazarus, E. M\"obius and M. Witte (1996), Ionization processes in the heliosphere - rates and methods of their determination, {\tenit Space Sci. Rev., 78\/}, 73-84.}

\hangindent=1.5pc \hangafter=1 \noindent
{Schmitt, Harrison H. (2006), Return to the Moon, Praxis Publishing Ltd.,
Copernicus Books, New York, p.~123.}

\hangindent=1.5pc \hangafter=1 \noindent
{Severny, A.B., E.I. Terez, and A.M. Zvereva (1974), Preliminary results obtained with an astrophotomter installed on Lunokhod 2, 
{\tenit Space Res. XIV\/}, 
Akademie-Verlag, 603-605.}

\hangindent=1.5pc \hangafter=1 \noindent
{Stern, S. Alan (1999), The lunar atmosphere: History, status, current problems, and context, {\tenit Rev. of Geophys., 37(4)\/}, 453-491.}

\hangindent=1.5pc \hangafter=1 \noindent
{Stubbs, T.J., R.R. Vondrak, and W.M. Farrell (2006), A dynamic fountain model
for lunar dust, {\tenit Adv. Space Res., 37(1)\/}, 59-66.}

\hangindent=1.5pc \hangafter=1 \noindent
{Stubbs, T.J., R.R. Vondrak, and W.M. Farrell (2007a), Impact of dust on lunar exploration, in {\tenit Dust in Planetary Systems\/}, (Workshop, September 26-30, 2005, Kauai, HI), edited by H. Kr\"uger and A.L. Graps, ESA Publications,
SP-643, pp. 239-244.} 

\hangindent=1.5pc \hangafter=1 \noindent
{Stubbs, T.J., J.S. Halekas, W.M. Farrell, and R.R. Vondrak (2007b), Lunar surface charging: A global perspective using Lunar Prospector data, in {\tenit Dust in Planetary Systems\/}, (Workshop, September 26-30, 2005, Kauai, HI), edited by H. Kr\"uger and A.L. Graps, ESA Publications,
SP-643, pp. 181-184.} 

\hangindent=1.5pc \hangafter=1 \noindent
{Stubbs, T.J., R.R. Vondrak, and W.M. Farrell (2007c), A dynamic fountain model for dust in the lunar exosphere, in {\tenit Dust in Planetary Systems\/}, (Workshop, September 26-30, 2005, Kauai, HI), edited by H. Kr\"uger and A.L. Graps, ESA Publications,
SP-643, pp. 185-190.} 

\hangindent=1.5pc \hangafter=1 \noindent
{Stubbs, T.J., R.R. Vondrak, and W.M. Farrell, and M.R. Collier (2007d), Predictions of dust concentrations in the lunar exosphere, {\tenit (Chinese) J. of Astronautics\/}, 28, 166-167, 1000-1328(2007)Sup-0166-02.} 

\hangindent=1.5pc \hangafter=1 \noindent
{Taguchi, S., S.-H. Chen, M.R. Collier, T.E. Moore, M.C. Fok, K. Hosokawa and A. Nakao (2005), Monitoring the high-altitude cusp with low-energy neutral atom imager: Simultaneous observations from IMAGE and Polar, 
{\tenit J. Geophys. Res., 110\/}, 
A12204, doi:10.1029/2005JA011075.}

\hangindent=1.5pc \hangafter=1 \noindent
{Taguchi, S., K. Hosokawa, A. Nakao, M.R. Collier, T.E. Moore, A. Yamazaki, N. Sato, and A.S. Yukimatu (2006), Neutral atom emission in the direction of the high-latitude magnetopause for northward IMF: Simultaneous observations from IMAGE spacecraft and SuperDARN radar, {\tenit Geophys. Res. Lett., 33\/}, 
L03101, doi:10.1029/2005GL025020.}

\hangindent=1.5pc \hangafter=1 \noindent
{Taylor, L.A. (2007), Everything you ever wanted to know about lunar dust, {\tenit Workshop on Science Associated with the Lunar Exploration Architecture\/}, 
NASA Advisory Council, Tempe, AZ.}

\hangindent=1.5pc \hangafter=1 \noindent
{Vondrak, R.R. (2007), Effects of ISRU on the lunar environment, {\tenit Workshop on Science Associated with the Lunar Exploration Architecture\/}, 
NASA Advisory Council, Tempe, AZ.}

\hangindent=1.5pc \hangafter=1 \noindent
{Wang, X., M. Horanyi, Z. Sternovsky, S. Robertson and G.E. Morfill (2007), A laboratory model of the lunar surface potential near boundaries between sunlit and shadowed regions, {\tenit Geophys. Res. Lett., 34\/}, doi:10.1029/2007GL030766, L16104.}

\hangindent=1.5pc \hangafter=1 \noindent
{Wimmer-Schweingruber, Robert F. and Peter Bochsler (2003), On the origin of inner-source pickup ions, {\tenit Geophys. Res. Lett., 30(2)\/}, doi:10.1029/2002GL015218, 49-1-49-4.}

\hangindent=1.5pc \hangafter=1 \noindent
{Zook, H.A. and J.E. McCoy (1991), Large scale lunar horizon glow and a high altitude lunar dust exosphere, {\tenit Geophys. Res. Lett., 18(11)\/}, 2117-2120.}

\vfill \eject

\noindent {\tenbf Figure Captions}

\nobreak

\vskip 5pt

\nobreak

\noindent
{\tenbf Figure 1.} The dust concentration in m${}^{-3}$ as a function of
altitude above the lunar surface at the terminator for dust grain radii from 0.1-6~$\mu$m (solid line) from Stubbs et al. [2007d] (based on a numerical integration of the Murphy and Vondrak [1993] model) and the approximation (dashed line) given by equation~(16).
\vskip 0.10in

\noindent
{\tenbf Figure 2.} The range of hydrogen through silicon as a function of hydrogen speed over the typical range of solar wind speeds. Note that even at the highest solar wind speeds, $\sim$1000~km/s, solar wind protons will not all penetrate spherical dust grains with radius 0.1~$\mu$m. However, at solar wind speeds slightly greater than average ($\sim$440~km/s), solar wind protons will completely penetrate 0.01~$\mu$m grains.
\vskip 0.10in


\noindent
{\tenbf Figure 3.} Geometry of neutral solar wind viewing at the lunar terminator. The Sun is to the left and the solar wind flows antisunward with speed $v$. The lunar radius, 1738~km, is designated $R$. A point $x$ along the line-of-sight corresponds to an altitude $z$ above the lunar surface.
\vskip 0.10in

\noindent
{\tenbf Figure 4.} Prediction for the ratio of neutral solar wind flux to ionized solar wind flux considering only the lunar dust population at the terminator with grain radii in excess of 0.1~$\mu$m (left axis, solid line). Also, the neutral solar wind count rate that would be observed by an instrument like IMAGE/LENA assuming a solar wind flux of $3{\rm x}10^8$/cm${}^2$/s and an efficiency of $2{\rm x}10^{-4}$ as a function of solar wind speed (right axis, dashed line).
\vskip 0.10in

\noindent
{\tenbf Figure 5.} The correction factor $C(a_{\rm tr})$ as a function of 
${a_{\rm tr}/a_{\rm min}}$. $C(a_{\rm tr})$ modifies the neutral solar wind flux calculation to take into account the effect of the solar wind fully penetrating the dust grain. Recall that $a_{\rm tr}$ is a function of solar wind speed.
\vskip 0.10in

\noindent
{\tenbf Figure 6.} Prediction for the ratio of neutral solar wind flux to ionized solar wind flux considering the lunar dust population at the terminator with grain radii in excess of 0.01~$\mu$m (left axis, solid line). Also, the neutral solar wind count rate that would be observed by an instrument like IMAGE/LENA instrument assuming a solar wind flux of $3{\rm x}10^8$/cm${}^2$/s and an efficiency of $2{\rm x}10^{-4}$ as a function of solar wind speed (right axis, dashed line).
\vskip 0.10in

\vfill \eject
\end